\documentclass[preprint,12pt]{elsarticle/elsarticle}
\pdfoutput=1
\usepackage{amsmath,amssymb,graphicx,subfig,bm,euscript,cancel,mathtools,setspace,color,mathrsfs,hyperref}
\usepackage{geometry}
 \def\eq#1{Eq.~(\ref{#1})}

 \def\Fig#1{Fig.~{\ref{#1}}}
 \def\beq{\begin{equation}}
 \def\eeq{\end{equation}}
 \def\beqa{\begin{eqnarray}}
 \def\eeqa{\end{eqnarray}}
 
 \newcommand{\bs}{\boldsymbol}
 \def\one{\!\!{\hbox{ 1\kern-.8mm l}}}

 \newcommand{\ex}[1]{{\rm e}^{#1}}
 \renewcommand{\d}{{\rm d}}
 \def\ii{{\rm i}}

\makeatletter
\newcommand{\dotminus}{\mathbin{\text{\@dotminus}}}
\newcommand{\@dotminus}{%
  \ooalign{\hidewidth\raise1ex\hbox{.}\hidewidth\cr$\m@th-$\cr}%
}
\makeatother

\DeclareMathAlphabet\EuRoman{U}{eur}{m}{n}
\SetMathAlphabet\EuRoman{bold}{U}{eur}{b}{n}

\newcommand{\ie}{\emph{i.e.}}

\newcommand{\osp}{\ensuremath{\text{OSp}(1|2)}}
\newcommand{\ve}{\varepsilon}

\allowdisplaybreaks
\begin{document}
\begin{frontmatter}
\journal{Physics Letters B}
\title{
Schwinger-type parametrization of open string worldsheets}
\author{Sam Playle\corref{cor1}}
\cortext[cor1]{Corresponding author}
\ead{playle@to.infn.it}
\author{Stefano Sciuto\corref{}}
\ead{sciuto@to.infn.it}
\address{{\sl Dipartimento di Fisica, Universit\`a di Torino  \\
and INFN, Sezione di Torino}\\
{\sl Via P. Giuria 1, I-10125 Torino, Italy}}
\begin{abstract}
\noindent%
A parametrization of (super) moduli space near the corners corresponding to bosonic or Neveu-Schwarz open string degenerations is introduced for worldsheets of arbitrary topology. With this parametrization, Feynman graph polynomials arise as the $\alpha ' \to 0$ limit of objects on moduli space. Furthermore, the integration measures of string theory take on a very simple and elegant form.
\end{abstract}
\end{frontmatter}
\section{Introduction}
\label{intro}
A very useful parametrization of the moduli of multiloop Riemann surfaces is given by Schottky groups, which manifested themselves automatically in the earliest approaches to multiloop string amplitudes. 
%
In this letter we describe (in section \ref{param}) a scheme for co-ordinatizing the moduli space of orientable \emph{open string} worldsheets, in which all $3g-3+n$ real moduli are realized as `lengths' of plumbing fixtures. In section \ref{qftlim} we see how Feynman graphs with various distinct topologies arise as the $\alpha ' \to 0$ limit of such worldsheets, given an appropriate mapping between dimensionless pinching parameters $p_i$ and Schwinger parameters $t_i$. In section \ref{NS} we show how the construction can be extended to the Neveu-Schwarz sector of superstrings, and present the elegant form taken by the leading part of the string measure in the pinching moduli.
 The pinching moduli are `canonical parameters' in the sense of section 6.3 of reference \cite{Witten:2012ga}, so their use makes Berezin integration on supermoduli space unambiguous.
 Proofs omitted in this letter are to be provided in a forthcoming work \cite{forthcoming}.

\section{The parametrization}
\label{param}
We are interested in describing worldsheets near complete ``open string'' degenerations; in such regions of moduli space the worldsheets may be constructed from 3-punctured discs glued together with strips. The topologically distinct degenerations can be classified as cubic ribbon graphs (\ie~graphs with a fixed cyclic ordering of the three edges incident on each vertex). Given a (not necessarily planar) cubic ribbon graph, we want to find ``pinching parameters'' $\{p_i\}$, \ie~local coordinates on Schottky space such that taking $p_i \to 0$ gives the corner corresponding to that degeneration.

To achieve this, we will provide an algorithm for writing down the $g$ Schottky group generators $\gamma_i$ and the $n$ positions of punctures $x_j$ as functions of the pinching parameters for a given cubic ribbon graph.

The algorithm may be arrived at by considering transition functions on a surface obtained by gluing together 3-punctured discs with open-string plumbing fixtures. All transition functions will be composed of two fundamental ones: one that cycles between local coordinates around the three punctures on a disc, and one which moves from one end of a plumbing fixture to the other.

Let us consider first of all a 3-punctured disc. Let the punctures be labelled $a_1$, $a_2$ and $a_3$ with a clockwise ordering. We will need three local coordinate charts $z_1$, $z_2$, $z_3$ which vanish at their respective punctures; $z_i(a_i) =0$. The upper-half-plane is the image of the disc under $z_i$ and its boundary is mapped onto the projective real line. Let us also specify
\begin{align}
z_i(a_{i+1}) & = \infty \, \, ;
&
z_i(a_{i-1}) & = 1 \, \, ,
\end{align}
where the indices are mod 3. Then there is a unique M\"obius map $\rho$ which acts as a \textbf{transition function cycling the three charts}. We want to have $z_i = \rho(z_{i+1})$, then we need
\begin{align}
\rho(0) & = \infty \, \, ;
&
\rho(\infty) & = 1 \, \, ;
&
\rho(1) & = 0 \, \, .
\end{align}
This is given by $\rho(z) = 1 - 1/z$, or as a matrix acting on the homogeneous coordinates,
\begin{align}
\rho & = \left(\!\!\begin{array}{cc} 1 & -1 \\ 1 & 0 \end{array}\!\!\right) \, . \label{rhodef}
\end{align}
which, of course, satisfies $\rho^3 = \text{Id}$.
So in general, on a 3-punctured disc the transition functions between these canonical charts are given by
\begin{align}\begin{cases}
\rho & \leftrightarrow  \text{move anticlockwise around the disc }\\
\rho^{-1} & \leftrightarrow \text{move clockwise around the disc. }
\end{cases}
\end{align}

The other ingredient is the \textbf{open string plumbing fixture}. Suppose our surface includes two charts $z$, $w$ whose images are contained in the upper-half-plane and include semi-discs of radius 1 centred on 0. Then if we fix a ``pinching parameter'' $p$ with $0<p<1$ and cut out the semi-discs $|z|<p$, $|w|<p$ we can impose the equation
\begin{align}
z\,w & = - p \, \, , \label{plumbingfixture}
\end{align}
for $|z|<1$, $|w|<1$, which we call an open string plumbing fixture between the two charts. Topologically, the effect is to attach a strip to the boundary of the surface, either adding a `handle' or joining two previously disconnected components. When we take $p\to 0$ the strip degenerates leaving a node joining $z^{-1}(0)$ to $w^{-1}(0)$.

We can view the plumbing fixture as a transition function from the chart at one end to the chart at the other: let us define a M\"obius map $\sigma_p$ such that \eq{plumbingfixture} can be written as $w = \sigma_p(z)$, \ie~$\sigma_p(z) \equiv - p/z$, or as a matrix
\begin{align}
\sigma_p & = \frac{1}{\sqrt{p}} \left(\!\!\begin{array}{cc} 0 & - p \\ 1 & 0 \end{array}\!\!\right) \, .
\end{align}
We can summarize its use as
\begin{align}
\sigma_p & \leftrightarrow \text{ traverse a plumbing fixture with pinching parameter }p\, \, .
\end{align}

Now let us consider a cubic ribbon graph $\Gamma$. Let us assign \emph{three} coordinate charts to each vertex, with one associated to each incident half-edge. We can write down a sequence composed of the following two moves taking us from one chart to any other one:
\begin{itemize}
\item Moving (anti)clockwise between two charts associated to different half-edges incident at the same vertex.
\item Moving from a chart associated to a half-edge of an internal edge $E_k$ to a chart associated to its half-edge at the vertex at the other end.
\end{itemize}
It's crucial that an internal edge not be traversed before first moving onto the chart associated to its half-edge.

A sequence of such moves can be translated into a transition function with the following dictionary:
\begin{align}
\begin{cases}
 \text{move anticlockwise around a vertex}& \leftrightarrow \rho \\
 \text{move clockwise around a vertex}& \leftrightarrow \rho^{-1} \\
 \text{traverse }E_k & \leftrightarrow \sigma_{p_k} \equiv \sigma_k \, ,
\end{cases} \label{dictionary}
\end{align}
where we have associated a pinching parameter $p_k$ to every internal edge $E_k$.

Note that for multiply-connected graphs, this procedure gives multiple, distinct transition functions from one chart to another, since there are multiple paths between each pair of charts and each path gives a different transition function. This is because there is a Schottky group: each transition function is well-defined modulo the group action.

To be more explicit, let us pick a \textbf{``base chart''} $z$ (\ie~a choice of one of the vertices in $\Gamma$ \emph{and} one of its incident half-edges). If the surface has $g$ loops, then we can find $g$ homologically independent closed paths $P_i$ starting and ending at $z$. For each closed path, we can use \eq{dictionary} to write down a M\"obius map; these $g$ M\"obius maps are the Schottky group generators $\gamma_i$.

Furthermore, suppose $\Gamma$ has $n$ external edges (corresponding to punctures in the surface). Each external edge has a coordinate chart, in which the punctures are at 0. We can write down paths $P_j$ from these charts back to the base chart $z$, and again using the dictionary \eq{dictionary}, we can find M\"obius maps $V_j$ which are transition functions from these charts to the base chart $z$. Then the positions of the punctures as seen in the base chart will be given as
\begin{align}
x_j & = V_j(0) \, .
\end{align}

So we have defined a Riemann surface by a set of transition functions which depend on a set of parameters $\{p_i\}$. The number of parameters equals the number of internal edges, which by elementary graph topology is $3g-3+n$, coinciding with the real dimension of open string Schottky space. The canonical Schottky coordinates (multipliers, fixed points and puncture coordinates) can be expressed in terms of the $p_i$'s, so these provide a new set of coordinates for Schottky space. Moreoever, in the limit $p_i \to 0$, the surface totally degenerates into a collection of 3-punctured discs joined together at nodes, with the topology corresponding to the graph $\Gamma$ used to define the $p_i$'s.

Use of these pinching parameters as integration variables on moduli space gives the bosonic string measure an elegant and simple form which we describe at the end of section \ref{SuperstringMeasure}, since it is similar to the analogous expression for the NS sector of superstrings \cite{forthcoming}.
\subsection{An example at 3-loop}
\label{mercsec}
\begin{figure}
\centering
\subfloat[]{ \includegraphics[scale=1.2]{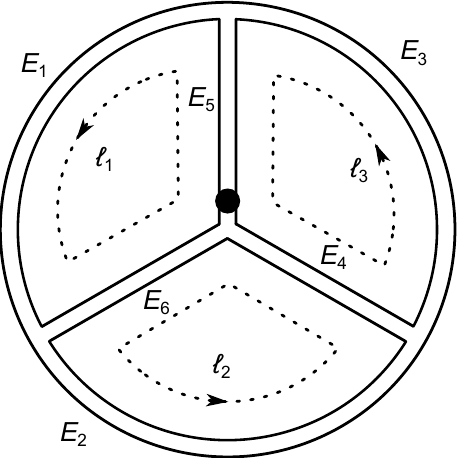} \label{fig:1a} }
\subfloat[]{ \includegraphics[scale=1.2]{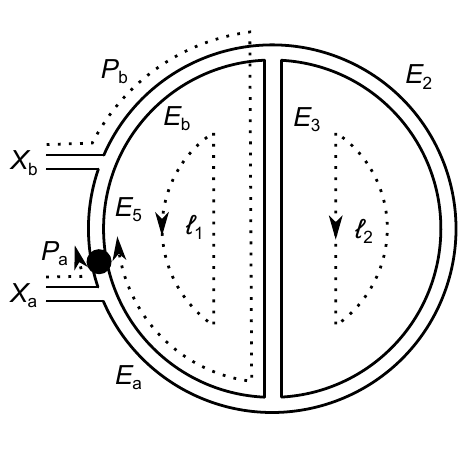} \label{fig:1b} }
\caption{Two cubic ribbon graphs. Internal edges are labelled by $E_i$ and external edges are labelled with $X_j$. The big dots indicate the chosen base charts. Dotted lines indicate paths: loops are labelled $\ell_k$ while paths from $X_j$ to the base chart are labelled $P_j$.}\label{fig1}
\end{figure}
Let us consider the 3-loop ``Mercedes-Benz'' diagram shown in \Fig{fig:1a}. We can write down the Schottky group for this graph according to the procedure in section \ref{param}. The big dot indicates a choice of coordinate chart to use as our base chart. A basis of three loops $\ell_1$, $\ell_2$, $\ell_3$ is indicated. For each of these loops, we can write down a sequence of the basic `moves' needed to go around the loop and arrive back in the base chart, as follows (reading right-to-left):
\newcommand{\cw}{\textsc{cw}}
\newcommand{\acw}{\textsc{acw}}
\begin{align}
\ell_1 & =  \cw \, \cdot \, E_6 \, \cdot \, \cw \, \cdot \, E_1 \, \cdot \, \cw \, \cdot \, E_5 \\
\ell_2 & = \acw \, \cdot \, E_4 \, \cdot \,  \cw \, \cdot \,  E_2\, \cdot \, \cw\, \cdot \, E_6 \, \cdot \, \acw \\
\ell_3 & =  E_5 \, \cdot \, \cw \, \cdot \,  E_3 \, \cdot \,  \cw \, \cdot \,  E_4 \, \cdot \,\cw \, ,
\end{align}
where \textsc{(a)cw} means ``move to the chart that is (anti)clockwise from the current one on the same vertex'' and $E_i$ means ``move to the chart at the other end of the edge $E_i$''. Then the dictionary \eq{dictionary} gives us the following matrices as the three Schottky generators:
\begin{align}
\gamma_1 & = \rho^{-1} \, \sigma_{6} \, \rho^{-1} \, \sigma_1 \, \rho^{-1} \, \sigma_5 = \frac{1}{\sqrt{k_1}} \left(\!\!\begin{array}{cc} 1 & 0 \\ 1 + p_6(1+p_1) & k_1 \end{array}\!\!\right) \\
\gamma_2 & = \rho \, \sigma_4 \, \rho^{-1} \, \sigma_2 \, \rho^{-1} \, \sigma_6 \, \rho  = \frac{-1}{\sqrt{k_2}}\left(\!\! \begin{array}{cc}
1+p_4(1+p_2(1+p_6)) & -1 - p_4(1+p_2) \\ p_4(1+p_2(1+p_6)) & -p_4(1+p_2)
\end{array}\!\!\right) \\
\gamma_3 & = \sigma_5 \, \rho^{-1} \, \sigma_3 \, \rho^{-1} \, \sigma_4 \, \rho^{-1} = \frac{1}{\sqrt{k_3}} \left(\!\! \begin{array}{cc}
k_3 & - p_5(1+p_3(1+p_4)) \\ 0 & 1
\end{array}\!\!\right)
\end{align}
where the multipliers $k_i$ are
\begin{align}
k_1 & = p_1 \, p_5 \, p_6
&
k_2 & = p_2 \, p_4\, p_6
&
k_3 & = p_3 \, p_4 \, p_5 \, ,
\end{align}
\ie~simply the products of the pinching parameters of the edges in the respective loops (this is true in general whenever a loop $\ell_i$ is conjugate to one whose turns are either all \cw~or all \acw).
The attractive and repulsive Schottky fixed points, $u_i$ and $v_i$ respectively, can be computed as
\begin{align}
u_1 & = 0
&
v_1 & = \infty
&
v_2 & = 1
&
v_3 & = \frac{1 + p_6(1+p_1)}{p_6(1+p_1(1+p_5))}
\end{align}
\begin{align}
u_2 & = \frac{1+p_6(
(1+p_4 + p_1(1+p_4(1+p_2)(1+p_5)))
+p_1\,p_2\,p_4\,p_5\,p_6)}
{p_6(1+p_4(1+p_2))(1+p_1(1+p_5))
} \shortintertext{and}
 u_3 & = \frac{1 + p_5((1 + p_6+p_3(1+p_6(1+p_1)(1+p_4)))+p_1\,p_3\,p_4\,p_5\,p_6)}{p_5\,p_6(1+p_3(1+p_4))(1+p_1(1+p_5))}
\, .
\end{align}
\subsection{An example with external edges}
\label{g2n2}
For a second example, let us consider the $g=2$, $n=2$ graph in \Fig{fig:1b}. The two loops may be written as
\begin{align}
\ell_1 & = E_5 \, \cdot \, \cw \, \cdot \, E_b \, \cdot \, \cw \, \cdot \, E_3 \, \cdot \, \cw \, \cdot \, E_a \, \cdot \, \cw \\
\ell_2 & = \acw \, \cdot \,E_a \, \cdot \, \acw \, \cdot \, E_3 \, \cdot \,\cw \, \cdot \, E_2\, \cdot \, \acw \, \cdot \,E_a\, \cdot \, \cw
\end{align}
so the Schottky group generators are
\begin{align}
\gamma_1 & = \sigma_5 \, \rho^{-1} \, \sigma_b \, \rho^{-1} \, \sigma_3 \, \rho^{-1}\,  \sigma_a \, \rho^{-1}  \\
\gamma_2 & = (\rho\, \sigma_a \, \rho) \, \sigma_3 \, \rho^{-1} \, \sigma_2 \, \rho^{-1} \, (\rho\, \sigma_a \, \rho) ^{-1} \, ,
\end{align}
whose fixed points and multipliers may be computed straightforwardly.
The paths $P_i$ from the external edges $X_i$ to the base chart may be written as
\begin{align}
P_a & = \cw & P_b & = \acw \, \cdot \, E_a \, \cdot \, \acw \, \cdot \, E_3 \, \cdot \, \acw \, \cdot \, E_b  \, \cdot \, \cw \label{g2n2Ps}
\shortintertext{so}
V_a & = \rho^{-1} & V_b  &   = \rho \, \sigma_a \, \rho \, \sigma_3 \, \rho \, \sigma_b \, \rho^{-1} \label{g2n2Vs}
\end{align}
hence the coordinates of the punctures in the base chart are
\begin{align}
x_a & = V_a(0) = 1 \, , &
x_b & = V_b(0) = \frac{1 + p_b(1+p_3(1+p_a))}{p_a\,p_b\,p_3} \, .
\end{align}
\section{The field theory limit}
\label{qftlim}
 Suppose a cubic ribbon graph $\Gamma$ is used to parametrize a Schottky group according to the procedure in section \ref{param}. Let us hypothesize the following expression for the $3g-3+n$ pinching parameters $p_i$ in terms of Schwinger parameters $t_i$ (where $\alpha ' $ is the Regge slope):
\begin{align}
p_i & = \ex{ - t_i / \alpha ' } \, . \label{ptmap}
\end{align}
With this, we can study the $\alpha ' \to 0$ asymptotics of various objects defined in terms of the open string worldsheets. We find that the limiting behaviour is given in terms of purely graph-theoretic objects defined in terms of $\Gamma$, where the $t_i$ are taken as Schwinger parameters for the corresponding internal edges $E_i$.
\subsection{Period matrix}
\label{permat}
The period matrix of a Riemann surface given by a Schottky group (with a compatible marking) is equal to the following series \cite{Mandelstam:1985ww}
\begin{align}
\tau_{ij} & = \frac{1}{2 \pi \ii } \Big( \delta_{ij} \, \log k_i \, - \, {}^{(i)} {\sum_{\gamma_\alpha}}'{}^{(j)} \log \frac{u_i - \gamma_\alpha(v_j)}{u_i - \gamma_\alpha(u_j)} \frac{v_i - \gamma_\alpha(u_j)}{v_i - \gamma_\alpha(v_j)} \Big) \label{pmseries}
\end{align}
where the summation is over all Schottky group elements $\gamma_\alpha$ whose left-most factor is not $\gamma_i^{\pm n}$ and whose right-most factor is not $\gamma_j^{\pm n}$. We can compute this, for example, for the 3-loop worldsheet described in section \ref{mercsec}. We find
\begin{align}
\tau_{ij} & = \frac{1}{2 \pi \ii } \left( \!\! \begin{array}{ccc} \log p_1  p_5 p_6  & - \log p_6 & - \log p_5 \\
- \log p_6 & \log p_2 p_4 p_6 & - \log p_4 \\
- \log p_5 & - \log p_4 & \log p_3 p_4 p_5
\end{array}\right) \, \, + \, \, {\cal O}( p_i) \, . \label{mercpm}
\end{align}
All Schotty group elements other than the identity give an ${\cal O}(p_i)$ contribution in \eq{mercpm}.

We can also compute the \emph{graph period matrix} for a $g$-loop graph $\Gamma$ with a basis $\{\ell_i\}$ of loops. This is given by \cite{Dai:2006vj}
\begin{align}
\theta_{ij} & = \sum_k \langle \ell_i , \ell_j \rangle ^k \, t_k\, ,
\end{align}
where for paths $P_1$, $P_2$ we define
 \begin{align}
\langle P_1, P_2  \rangle^k  & \equiv \begin{cases}
1 & \text{ if }P_1 \text{ and }  P_2 \text{ cross }E_k \text{ in the same direction}\\
- 1 & \text{ if }P_1 \text{ and } P_2 \text{ cross }E_k \text{ in opposite directions}\\
0 & \text{ if }P_1 \text{ and } P_2\text{ do not both cross }E_k.
\end{cases} \label{angdef}
\end{align}
The graph period matrix for the graph in \Fig{fig:1a} is given by
\begin{align}
\theta_{ij} & = \left( \!\! \begin{array}{ccc}
t_1 + t_5 + t_6 & - t_6 & - t_5 \\
- t_6 & t_2 + t_4 + t_6 & - t_4 \\
- t_5 & - t_4 & t_3 + t_4 + t_5
\end{array}\right) \, . \label{mercgraphpm}
\end{align}
Clearly, with the use of \eq{ptmap} the matrices in \eq{mercpm} and \eq{mercgraphpm} satisfy
\begin{align}
\ii\,  \theta_{ij} & =\lim_{\alpha ' \to 0}  2 \pi \alpha' \tau_{ij} \, . \label{pmlimit}
\end{align}
In fact, \eq{pmlimit} is true in general as a relation between an arbitrary $g$-loop, $n$-point cubic graph $\Gamma$ and the surface parametrized with it as in section \ref{param} \cite{forthcoming}.

The determinant of a graph's period matrix is the graph's first Symanzik polynomial, which appears in the denominator of the corresponding Feynman integrals \cite{Vanhove:2014wqa}. \eq{pmlimit} elucidates how this can arise from the powers of $\det( 2 \pi\,\text{Im}\, \tau)$ which appear in the denominator of the string measure on moduli space.
\subsection{Green's function}
This parametrization also allows us to see how the worldline Green's function can arise as the $\alpha ' \to 0$ limit of the worldsheet Green's function.

We use the results of \cite{Dai:2006vj}, where it is shown that the worldline Green's function between two external edges $X_1$, $X_2$ on a $g$-loop graph $\Gamma$ with a basis $\{\ell_i\}$ of loops may be written as
\begin{align}
G_{X_1X_2} & = - \frac{1}{2} s \, + \, \frac{1}{2} \, \vec v \, \cdot\, \theta^{-1} \cdot \, \vec v \, .
\end{align}
where we've picked some path $P$ from $X_1$ to $X_2$ and then in terms of \eq{angdef},
\begin{align}
s & \equiv \langle P , P \rangle
&
v_i & \equiv \langle \ell_i , P \rangle \, . \label{svdef}
\end{align}
$G$ can be found from the $\alpha' \to 0$ limit of
\begin{align}
\widehat{\cal G}(x_1,x_2) & \equiv {\cal G}(x_1 , x_2) - \frac{1}{2} \log (V_1'(0) V_2'(0))
\end{align}
where $V_i$ is the transition function that goes from the chart associated with the external edge $X_i$ to the base chart $z$, and $x_i = V_i(0)$ is the $z$ coordinate of the puncture. ${\cal G}(w,z)$ is the worldsheet Green's function given by \cite{DiVecchia:1996uq}
\begin{align}
{\cal G}(z,w) & = \log E(z,w) - \frac{1}{2} \Big( \int_z^w \vec\omega \Big) \cdot ( 2 \pi \text{Im}\, \tau)^{-1} \cdot \Big( \int_z^w \vec\omega \Big) \, .
\end{align}
Here $E(z,w)$ is the Schottky-Klein prime form
\begin{align}
E(z,w) & = (z-w){\prod_{\alpha}}' \frac{ z- \gamma_\alpha(w) }{z - \gamma_\alpha(z)}\frac{w - \gamma_\alpha(z)}{w - \gamma_\alpha(w)}
\end{align}
where the Schottky group product includes one from each pair of inverse elements $\{\gamma_\alpha, \gamma_\alpha^{-1}\}$. $\omega_i$ are the Abelian differentials, given by
\begin{align}
\omega_i (z) & = {\sum_\alpha}{}^{(i)} \Big( \frac{1}{z - \gamma_\alpha(u_i)} - \frac{1}{z - \gamma_\alpha(v_i)} \Big) \, \d z \,
\end{align}
where the sum is over all Schottky group elements whose right-most factor is not $\gamma_i^{\pm n}$. We find that
\begin{align}
\frac{1}{2}\, s & = \lim_{\alpha ' \to 0} \alpha ' \log \frac{E(x_1,x_2)}{\sqrt{V_1'(0)V_2'(0)}}
&
v_i & = - \lim_{\alpha ' \to 0} \alpha' \int_{x_1}^{x_2} \omega_i \,
\end{align}
which along with \eq{pmlimit} gives
\begin{align}
G_{X_1X_2} & =- \lim_{\alpha ' \to 0 } \alpha'\,  \widehat{\cal G}(x_1, x_2) \, .
\end{align}
As an example, consider the $g=2$, $n=2$ graph in \Fig{fig:1b}, whose corresponding pinching parametrization was worked out in section \ref{g2n2}.

Let us choose the path $P$  between the two external edges to be given in terms of \eq{g2n2Ps} by $P = P_b^{-1} \, \cdot \, P_a$; using \eq{svdef} this gives
\begin{align}
s & = t_a + t_b + t_3 \, &
\vec v & = (t_a + t_b + t_3 \, , \, - t_3 ) ^{\text t} \, .
\end{align}
From the Schottky group formulae, we can use
\begin{align}
V_a'(0) & = 1 \, , & V_b'(0)  &= \frac{1}{p_a\,p_b\,p_3} \, .
\end{align}
to find
\begin{align}
\log \frac{E(x_b, x_a)}{\sqrt{V_a'(0)V_b'(0)}} & = \log \frac{x_b- x_a}{\sqrt{V_a'(0)V_b'(0)}}+ {\cal O}(p_i) = - \frac{1}{2} \log(p_a\,p_b\,p_3)+ {\cal O}(p_i) \, , \label{slim}
\end{align}
which converges to $s/2 \alpha' $ after using \eq{ptmap} and taking $\alpha ' \to 0$.
Similarly, we can compute $\int_{x_a}^{x_b} \omega_i$. The only Schottky group element which contributes at leading order is the identity; we find
\begin{align}
\int_{x_b}^{x_a}\vec \omega(z) & = \Big[ \log \frac{z- u_i}{z - v_i} \Big]_{x_b}^{x_a} + {\cal O}(p_j) = \big( -\log(p_a\,p_b\,p_3) \, , \, \log(p_3) \big)^{\text{t}} + {\cal O}(p_j) \, ; \label{vlim}
\end{align}
again, this asymptotes to $\vec v / \alpha'$ in the limit $\alpha ' \to 0$. Thus, combining \eq{vlim} and \eq{slim} and computing the period matrix as in section \ref{permat}, we find that for the surface parametrized by the graph in \Fig{fig:1b},
\begin{align}
\lim_{\alpha' \to 0} \alpha' \widehat{\cal G}(x_b, x_a) & = \frac{t_5}{2} - \frac{t_5^2(t_2+t_3)}{2 \det \theta} = - G \, .
\end{align}
This holds in general. Since Feynman integrals for $\Phi^3$ scalar QFTs can be written down using only the worldline Green's function \cite{Frizzo:1999zx}, this clarifies how the Feynman diagrams arise from the various corners of moduli space in the corresponding string theory.
\section{Superstrings}
\label{NS}
The construction in section \ref{param} can be adapted for the Neveu-Schwarz (NS) sector of superstrings in the RNS formalism, in which the worldsheets are taken as super Riemann surfaces (SRS) \cite{Witten:2012ga}. We use the formalism of super Schottky groups \cite{Martinec:1986bq,Manin1986} following the notation of section 2.2 of \cite{Playle:2015sxa}.

A number of modifications must be made to the construction in section \ref{param}. Firstly, the 3-punctured discs must be replaced by SRS discs with three NS punctures (NNN discs). While 3-punctured discs have no moduli, NNN discs have one Grassmann-odd supermodulus. If the punctures are at $\bs a_1$, $\bs a_2$, $\bs a_3$ and $\bs z = z|\zeta$ is a global superconformal coordinate, then
\begin{align}
\Theta_{\bs a_1 \bs a_2 \bs a_3} & \equiv \pm \frac{\zeta_1 (\bs z_2 \dotminus \bs z_3) + \zeta_2(\bs z_3 \dotminus \bs z_1) + \zeta_3(\bs z_1 \dotminus \bs z_2) + \zeta_1 \zeta_2 \zeta_3}{\sqrt{(\bs z_1 \dotminus \bs z_2)(\bs z_2 \dotminus \bs z_3)(\bs z_3 \dotminus \bs z_1)} }\,
\end{align}
is a superprojective (pseudo)invariant, and a modulus of an NNN disc. Here $\bs z_i \equiv \bs z(\bs a_i)$ and $\bs z_i \dotminus \bs z_j \equiv z_i - z_j - \zeta_i \zeta_j$. To account for this, we attach an odd parameter $\Theta_i$ to every vertex $V_i$ in our cubic ribbon graph $\Gamma$. Whereas in the bosonic construction a single matrix $\rho$ is used to `rotate' around any vertex, now each vertex $V_i$ must get its own separate matrix $\bs \rho_{\Theta_i}$. It is given by the \osp~matrix
\begin{align}
\bs \rho_{\Theta} & \equiv \left( \begin{array}{cc|c} -1 & 1 & - \Theta \\ -1 & 0 & 0 \\ \hline - \Theta & 0 & 1 \end{array}\right)
\end{align}
which permutes the points with homogenous coordinates $(0,1|0)^{\text{t}}$, $(1,0|0)^{\text{t}}$, $(1,1|\Theta)^{\text{t}}$.

The second modification to the construction in section \ref{param} is that the internal edges of the cubic ribbon graph $\Gamma$ must be given orientations, \ie~$\Gamma$ must be a \emph{directed} cubic ribbon graph. The reason for this is that the NS plumbing fixture is asymmetric unlike its bosonic equivalent. If $\bs z = z|\zeta$ and $\bs w = w |\psi$ are two superconformal charts at opposite ends of an NS plumbing fixture, then they satisfy
\begin{align}
z \, w & = - \ve^2  \, ,
&
z \, \psi & = \ve \, \zeta \, ,
&
w \, \zeta & = - \ve \, \psi \, ,
&
\psi \, \zeta & = 0 \, , \label{NSplumb}
\end{align}
where we call $\ve$ the ``NS pinching parameter''.
\eq{NSplumb} is not symmetric under swapping $\bs z \leftrightarrow \bs w$ without also swapping $\ve \leftrightarrow - \ve$. We can rewrite \eq{NSplumb} in terms of a transition function as $\bs w = \bs \sigma_{\ve}(\bs z)$, where $\bs \sigma_{\ve}(\bs z) \equiv -\ve^2/z \, \big| \, \ve\, \zeta / z$, or as an \osp~matrix,
\begin{align}
\bs \sigma_{\ve} & \equiv \left( \begin{array}{cc|c} 0 & -\ve  & 0 \\ \ve^{-1} & 0 & 0 \\ \hline 0 & 0 & 1 \end{array}\right) \, .
\end{align}

The general idea of the approach to writing down a super-Schottky group from a directed cubic ribbon graph $\Gamma$ is the same as in the bosonic case. Each cubic vertex in $\Gamma$ has three associated superconformal charts (one for each incident half-edge). Let us pick one as a `base chart' $\bs z$. If $\Gamma$ is a $g$ loop graph, we can find $g$ independent closed paths starting at the base chart. After decomposing each path $\ell_i$ into the same set of basic `moves' as in the bosonic case, we can translate it into a super Schottky group generator $\bs \gamma_i$ using the following dictionary:
\begin{align}
\begin{cases}
 \text{move anticlockwise around the vertex }V_i& \leftrightarrow \bs \rho_{\Theta_i} \\
 \text{move clockwise around the vertex }V_i& \leftrightarrow \bs \rho_{\Theta_i}\!\!\!{}^{-1} \\
 \text{traverse }E_k\text{ in the marked direction} & \leftrightarrow \bs \sigma_{\ve_k} \\
 \text{traverse }E_k\text{ against the marked direction} & \leftrightarrow \bs \sigma_{\ve_k}\!\!\!{}^{-1} \, .
\end{cases} \label{NSdictionary}
\end{align}
Similarly, for each external edge $X_i$ in $\Gamma$, we find a path $P_i$ going from $X_i$ to the base chart and then use the dictionary \eq{NSdictionary} to translate $P_i$ into a transition function $\bs V_i$; the $\bs z$ coordinates of the corresponding NS puncture are $x_i |\xi_i = \bs V_i(0|0)$.

A $g$-loop cubic graph $\Gamma$ with $n$ external edges has $|V|=2g-2+n$ cubic vertices and $|E|=3g-3+n$ internal edges; since we associate a Grassmann odd parameter $\Theta_i$ to each vertex $V_i$ and a Grassmann-even parameter $\ve_k$ to each edge $E_k$, we see that we match the dimension of supermoduli space, $\dim(\mathfrak{M}_{g,n}) = 3g-3+n|2g-2+n$. For planar vacuum graphs with $n=0$ which have $|F|=g+1$ faces, we can use the supermoduli space dimension to check Euler's graph formula $|V|-|E|+|F|=2$.

Note that the parametrization chosen for $g=2$, $n=0$ in references \cite{Magnea:2013lna,Magnea:2015fsa} arises as a special case of the procedure described in this section. In those works, it was used to find the $\alpha ' \to 0$ limit of the NS superstring amplitude, correctly yielding the sum of 2-loop vacuum Feynman diagrams for the bosonic sector of ${\cal N}=4$ SYM in a particular gauge.
While the \emph{ad hoc} manipulation described in section 5 of \cite{Magnea:2013lna} was necessary to correctly choose which even supermoduli to fix before evaluating Berezin integrals (see section 3.4.1 of \cite{Witten:2012bh} for why this is important), and to rescale odd supermoduli to symmetrize the factors in the measure, the procedure described here prescribes the same outcome without ambiguity.

\subsection{The superstring measure}
\label{SuperstringMeasure}
The super Schottky group expression for the $g$-loop, $n$-point superstring measure in the NS sector is given by Eqs.~(30) and (31) of \cite{DiVecchia:1988jy}.
The leading holomorphic part (having dropped the period matrix determinants and nonzero mode parts of the functional determinants) is given in terms of canonical super Schottky variables by
\begin{align}
[ \d \bs m_0]_g^n & \equiv \frac{1}{ \d \bs V_{abc}} {  \prod_{i=1}^{n} \frac{ \d \bs x_i}{(D \bs V_i^{\zeta} )(0|0)}  } \prod_{j=1}^g \frac{ \d \bs u_i \, \d \bs v_j\,\d q_j }{\bs u_j \dotminus \bs v_j} \frac{(1 + q_j)^2 }{q_j^2} \label{NSschomeas} \, ,
\end{align}
and takes a simple and elegant  form in terms of these pinching parameters.
In \eq{NSschomeas} $\bs u_i$ and $\bs v_i$ are the attractive and repulsive fixed superpoints of $\bs \gamma_i$ and $q_i$ is its semimultiplier ($\bs \gamma_i$ is conjugate to $\bs z \mapsto q_i^2 z|q_i\zeta$). $\bs x_i = \bs V_i(0|0)$ is the position of the NS puncture associated to $X_i$, $ \bs V_i^\zeta$ is the odd part of $\bs V_i$ and $D$ is the superderivative. The superprojective volume element is given by
\begin{align}
\frac{1}{ \d \bs V_{abc}} & = \frac{\sqrt{(\bs a \dotminus \bs b)(\bs b \dotminus \bs c)(\bs c \dotminus \bs a)}}{\d \bs a \, \d \bs b \, \d \bs c} \, \d \Theta_{\bs a \bs b \bs c} , \label{superproj}
\end{align}
where $\bs a$, $\bs b$ and $\bs c$ are three superpoints chosen from among the $\bs x_i$, $\bs u_j$ and $\bs v_j$ to be gauge-fixed.

Expressed in terms of the pinching supermoduli, \eq{NSschomeas} takes on the following very simple form:
\begin{align}
[\d \bs m_0]_g^n & \propto \prod_{V_i}
\d \Theta_i\, \prod_{E_j} \,  \frac{\d \ve_j}{\ve_j^2} \, \prod_{B}  ( 1 +q_B) \, . \label{NSpimeas}
\end{align}
Here the product runs over all \emph{closed boundaries} $B$ of the ribbon graph $\Gamma$, meaning all closed paths in the graph whose decomposition involves either only \cw~or only \acw~turns at the vertices. $q_B$ is the semimultiplier of the super Schottky group element $\bs \gamma_B$ homologous to the path $B$, which is equal (modulo a sign) to the product of the NS pinching parameters $\ve_k$ of the edges $E_k$ in $B$. The product is also over all vertices $V_i$ and edges $E_j$ in the graph $\Gamma$.

The analogous part of the bosonic string theory measure on Schottky space also takes a simple form in terms of the pinching parameters, obtained from \eq{NSpimeas} by deleting the $\d \Theta_i$'s and replacing $\d \ve_j / \ve_j^2 \mapsto \d k_j / k_j^2$ and $ (1 + q_B) \mapsto ( 1- k_B)$ \cite{forthcoming}.
\subsection{Example}
\begin{figure}
\centering\includegraphics[scale=1.2]{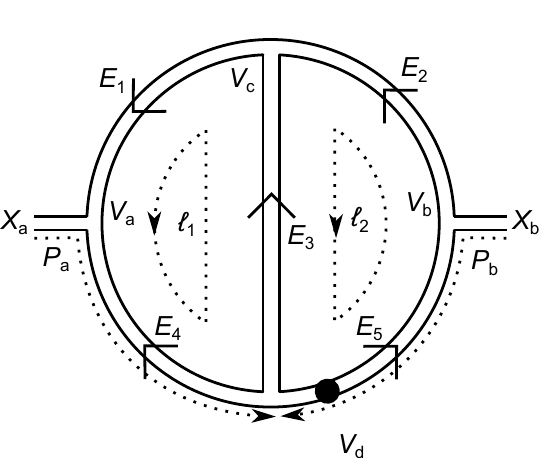}
\caption{A directed cubic ribbon graph with $g=2$, $n=2$. Vertices are labelled $V_i$ and internal edges have orientations marked by chevrons; the other labels have the same meanings as in \Fig{fig1}.}
 \label{fig2}
\end{figure}
Consider the $g=2$, $n=2$ graph shown in \Fig{fig2}. The loop basis indicated is given by
\begin{align}
\ell_1 & = \acw_d \, \cdot \, E_4^{-1}  \, \cdot \, \cw_a  \, \cdot \, E_1  \, \cdot \,\cw_c  \, \cdot \, E_3  \, \cdot \, \acw_d \, ,\\
\ell_2 & = \cw_d  \, \cdot \, E_3^{-1}  \, \cdot \, \cw_c  \, \cdot \, E_2 \, \cdot \, \cw_b  \, \cdot \, E_5 \, ,
\end{align}
so with the dictionary \eq{NSdictionary} we find that the super Schottky group generators are
\begin{align}
\bs \gamma_1 & = \bs \rho_d  \, \bs \sigma_4^{-1} \, \bs \rho_a^{-1} \, \bs \sigma_1 \, \bs \rho_c^{-1} \, \bs \sigma_3 \, \bs \rho_d \, , \\
\bs \gamma_2 & = \bs \rho_d^{-1} \, \bs \sigma_3^{-1} \, \bs \rho_c^{-1} \, \bs \sigma_2 \, \bs \rho_b^{-1} \, \bs \sigma_5 \, .
\end{align}
Their semimultipliers are given by
\begin{align}
q_1 & =  \ve_1 \, \ve_3 \, \ve_4 \, , &
q_2 & =\ve_2 \, \ve_3 \, \ve_5 \, .
\end{align}
The fixed points are $\bs u_i = \frac{U_i^1}{U_i^2}\big|\frac{U_i^3}{U_i^2}$ and $\bs v_i = \frac{V_i^1}{V_i^2}\big|\frac{V_i^3}{V_i^2}$ where $U_i$ and $V_i$ are eigenvectors satisfying $\bs \gamma_i U_i = q_i^{-1} U_i$ and $\bs \gamma_i V_i = q_i V_i$, given by
\begin{align}
U_1 & = \bs \rho_d^{-1} \, \hat U_1 \, ,
&
V_1 & = \bs \rho_d^{-1} \, \hat V_1 \, ,
&
U_2 & = \hat U_2 \, ,
&
V_2 & = \hat V_2 \, .
\end{align}
Here,
\begin{align}
\hat U_i & \equiv (1 - q_i^2 \, , \, A_{ I_i} \, | \,(1 + q_i) \Phi_{ I_i} )^{\text{t}} \, , &
\hat V_i & \equiv ( 0\, , \, 1 \,|\, 0 )^{\text{t}}
\end{align}
where $I_1  = (4,1,d,a,c)$, $I_2 = (3,2,d,c,b)$ and
\begin{align} \Phi_{ij\alpha\beta\gamma} & \equiv \Theta_\alpha  +\ve_i \Theta_\beta - \ve_i \ve_j \Theta_\gamma \, \\
A_{ij\alpha\beta\gamma} & \equiv 1 + \ve_i^2 + \ve_i^2 \ve_j^2 + \Theta_\alpha \, \Phi_{ij\alpha\beta\gamma} - \ve_i^2 \ve_j \Theta_\beta \Theta_ \gamma .
\end{align}
Similarly, the paths from the external edges $X_a$, $X_b$ to the marked base chart are given by
\begin{align}
P_a & = \acw_d  \, \cdot \, E_4^{-1}  \, \cdot \, \cw_a \, , &
P_b & = E_5^{-1}  \, \cdot \, \cw_b \, ,
\intertext{then with \eq{NSdictionary} we find}
\bs V_a & = \bs \rho_d \, \bs \sigma_4^{-1} \, \bs \rho_a \, , &
\bs V_b & = \bs \sigma_5^{-1} \, \bs \rho_b^{-1} \, ,
\end{align}
so $\bs x_a = \infty | 0 $ and $\bs x_b = - \ve_5^2 | - \ve_5\Theta_b$. Let's instead use $\bs V_{a(\epsilon)}(z|\zeta) \equiv \bs V_a (z+\epsilon|\zeta)$ to control the infinities until we take $\epsilon \to 0$ in the final result. The three gauge-fixed points are $(\bs a , \bs b , \bs c) = (\bs v_1 , \bs v_2 , \bs x_{a(\epsilon)})$, so the super-projective volume element \eq{superproj} is
\begin{align}
\frac{1}{\d V_{\bs v_1  \bs v_2 \bs x_{a(\epsilon)}}} & \sim \frac{1}{\epsilon \, \ve_4^{\,4}} \frac{\d \Theta_{\bs v_1  \bs v_2 \bs x_{a(\epsilon)}}}{\d \bs v_1\, \d  \bs v_2 \, \d \bs x_{a(\epsilon)}} \, ;
&
\d \Theta_{\bs v_1  \bs v_2 \bs x_{a(\epsilon)}} & \sim \d \Theta_d \, .
\end{align}
The denominators from the punctures are
\begin{align}
(D \bs V_{a(\epsilon)}^{\zeta} )(0|0) &= \frac{1}{\epsilon \, \ve_4} \, ; &
(D \bs V_{b}^{\zeta} )(0|0) & = \ve_5 \, .
\end{align}
The super-Jacobian of the change from the canonical super Schottky variables to the pinching parameters is given in block form as $\big(\begin{smallmatrix}A & |  B \\ \hline C & | D\end{smallmatrix}\big)$ where
\begin{align}
A & = \frac{\partial(u_1,u_2,q_1,q_2,x_b)}{\partial(\ve_1, \ldots, \ve_5)} & B & =  \frac{\partial(\theta_1,\theta_2,\xi_b,\Theta_{\bs v_1  \bs v_2 \bs x_{a(\epsilon)}})}{\partial(\ve_1, \ldots, \ve_5)}\\
C& =  \frac{\partial(u_1,u_2,q_1,q_2,x_b)}{\partial(\Theta_a, \ldots, \Theta_d)} & D & = \frac{\partial(\theta_1,\theta_2,\xi_b,\Theta_{\bs v_1  \bs v_2 \bs x_{a(\epsilon)}})}{\partial(\Theta_a, \ldots, \Theta_d)} \, ; \end{align}
its Berezinian is
\begin{align}
\frac{\det(A - B D^{-1} C)}{\det D } & = - 8 \frac{\ve_3^2 \ve_4 \ve_5}{(1+q_1)(1+q_2)} (\bs u_1 \dotminus \bs v_1)(\bs u_2 \dotminus \bs v_2)
\end{align}
which combines with the other factors in the measure \eq{NSschomeas} to give
\begin{align}
[ \d \bs m_0]_2^2 & =- 8\, \d \Theta_a\, \d \Theta_b\, \d \Theta_c\, \d \Theta_d\, \frac{\d \ve_1 \, \d \ve_2 \, \d \ve_3 \, \d \ve_4 \, \d \ve_5}{\ve_1^2 \, \ve_2^2\, \ve_3^2 \, \ve_4^2 \, \ve_5^2} (1 + q_1)(1+q_2) \, .
\end{align}
This is of the form \eq{NSpimeas} because modulo conjugation $\ell_1$ and $\ell_2$ are the two closed boundaries of \Fig{fig2} (the closed path which crosses the edges $E_5E_2E_1E_4^{-1}$ is not a closed boundary because of the external edges $X_a$ and $X_b$).
\section*{Acknowledgements}
The authors would like to thank R.~Russo and L.~Magnea for useful comments and suggestions and collaboration on related projects.  This work was supported by the Compagnia di San Paolo contract
``MAST: Modern Applications of String Theory'' \verb=TO-Call3-2012-0088=.
\section*{Bibliography}
\bibliographystyle{jhep}
\bibliography{rs}
\end{document}